\documentstyle[multicol,osa,epsf]{revtex}
\begin{document}
\title{Fully parallel algorithm for simulating dispersion-managed wavelength-division-multiplexed
optical fiber systems \footnote{\bf Accepted to Optics Letters (2001)} }

\author{P. M. Lushnikov$^{1,2}$
}

\address{$^1$ Theoretical Division, Los Alamos National Laboratory,
  MS-B284, Los Alamos, New Mexico, 87545
  \\
  $^2$ Landau Institute for Theoretical Physics, Kosygin St. 2,
  Moscow, 117334, Russia
  }


\maketitle

\begin{abstract}

An efficient numerical algorithm is presented for massively parallel simulations of
dispersion-managed wavelength-division-multiplexed optical fiber systems. The algorithm is based on
a weak nonlinearity approximation and independent parallel calculations of fast Fourier transforms
on multiple CPUs. The algorithm allows one to implement numerical simulations $M/2$ times faster
than a direct numerical simulation by a split-step method, where $M$ is a number of CPUs in
parallel network.
\end{abstract}
~~~~~~~ {\it OCIS codes:}  060.2330, 060.5530, 060.4370, 190.5530, 260.2030.
\\


A wavelength-division-multiplexed (WDM) dispersion-managed (DM) optical fiber system is the focus
of current research in high-bit-rate optical communications. High capacity of optical transmission
is achieved using both wavelength multiplexing and dispersion management. (See e.g. Ref.
\cite{99Gerges,00Mol2}). Wavelength multiplexing allows the simultaneous transmission of several
information channels, modulated at different wavelengths, through the same optical fiber.
 A dispersion-managed
\cite{kogelnik1,kurtzke1,chrapl1,smithknox1} optical fiber systems are designed to achieve low (or
even zero) path-averaged group-velocity dispersion (GVD) by periodically alternating the sign of
the dispersion along an optical fiber. This dramatically reduces pulse broadening. Second-order GVD
(dispersion slope) effects and  path-averaged GVD effects cause optical pulses in distinct WDM
channels to move with different group velocities. Consequently modeling of WDM systems requires
simulating a long time interval. Enormous computation resources are necessary to capture
accurately the nonlinear interactions between channels which deteriorates bit-rate capacity. The
large computational resources required  to simulate  WDM transmission over transoceanic distances
make parallel computation necessary. Here an efficient numerical algorithm is developed for
massive parallel computation of WDM systems. The required computational time is inversely
proportional to the number of parallel processors used. This makes feasible a full scale numerical
simulation of WDM systems on a workstation cluster with a few hundred processors.

Neglecting polarization effects and stimulated Raman scattering and Brillouin scattering, the
propagation of WDM optical pulses in a DM fiber is described by a scalar nonlinear Schr\"odinger
equation (NLS):
\begin{eqnarray}
i A_z -\frac{1}{2} \beta_2(z)  A_{tt}-\frac{i}{6}\beta_3(z)A_{ttt}   + \sigma(z) |A|^{2} u
\nonumber \\
=i\Big(-\gamma
+[\exp{(z_a\gamma)}-1]\Sigma^N_{k=1}\delta(z-z_k)\Big )A \nonumber \\
\equiv i G(z)A, \label{nls1}
\end{eqnarray}
where $z$ is the propagation distance along an optical fiber, $A(t,z)$ is the slow amplitude of
light; $\beta_2$ and $\beta_3$ are the first and second-order GVD respectively which are periodic
functions of $z$; $\sigma=(2\pi n_2)/(\lambda_0 A_{eff})$ is the nonlinear coefficient; $n_2$ is
the nonlinear refractive index; $\lambda_0=1.55\mu m$ is the carrier wavelength; $A_{eff}$ is the
effective fiber area; $z_k=kz_a\ (k=1,\ldots, N)$ are amplifier locations; $z_a$ is the amplifier
spacing; and $\gamma$ is the loss coefficient. Note that distributed amplification can be also
included in $G(z)$ without changing the following analysis.

The change of variables $u=A e^{ -\int^z_0 G(z')dz'}$  results in the NLS with the $z$-dependent
nonlinear coefficient $c(z)\equiv \sigma(z)\exp{\Big(2\int^z_0G(z')dz'\Big)}$:
\begin{equation}
i u_z -\frac{1}{2} \beta_2(z) u_{tt}-\frac{i}{6}\beta_3(z)u_{ttt}  +  c(z)|u|^{2} u =0.
\label{nls0}
\end{equation}

By applying Fourier transform $\hat u(\omega,z)=\int^\infty_{-\infty}u(t,z)e^{\imath \omega t}dt$
to Eq. $(\ref{nls0})$, changing variables $\hat u(\omega,z) \equiv \hat \psi(\omega,z)
\exp{\Big(\frac{i}{2}\int^{z}_{0}dz' [\omega^2
 \beta_2(z')+\frac{\omega^3}{3}\beta_3(z')]  \Big)}$
and integrating Eq. $(\ref{nls0})$ over $z$ from $z_0$ to $z$ one obtains the following
integro-differential equation:
\begin{eqnarray} \label{nlsomega0}
 \hat \psi\big (\omega,z\big)=\hat \psi(\omega,z_0)+
 \nonumber \\
 i R\big(\hat \psi[\omega,z],\omega,z,z_0\big),
\end{eqnarray}
where
%
\begin{eqnarray} \label{Rdef2}
R\big(\hat v[\omega,z],\omega,z,z_0\big)= \frac{1}{(2\pi)^2}\int d\omega_1 d\omega_2d\omega_3
\int^{z}_{z_0}dz'
 \nonumber \\
\times \, c(z')\, \hat v^{(z')}(\omega_1,z')  \hat v^{(z')}(\omega_2, z')  \hat
v^{*\,(z')}(\omega_3,z')
\nonumber \\
\times
  \exp{\Big( -\frac{i}{2}\int^{z'}_{0}dz'' [\omega^2
 \beta_2(z'')+\frac{\omega^3}{3}\beta_3(z'')]  \Big)}
 \nonumber
\\
\times \delta(\omega_1+\omega_2-\omega-\omega_3),
\nonumber \\
\hat v^{(z)}(\omega,z) \equiv \hat v(\omega, z) \nonumber
\\
\times \exp{\Big(\frac{i}{2}\int^{z}_{0}dz' [\omega^2
 \beta_2(z')+\frac{\omega^3}{3}\beta_3(z')]  \Big)}.
\end{eqnarray}

If the nonlinearity is small: $z_{nl}\gg z_{disp}$, where
 $z_{nl}\equiv 1/|p|^2$ is a characteristic nonlinear length,
 $z_{disp}\equiv\tau^2/|\beta_2|$ is the dispersion length; $p$ and $\tau$ are  typical pulse amplitude and
width respectively. Then one can conclude that $\hat \psi(\omega, z)$ is a slow function of $z$ on
any scale $L\ll z_{nl}$ because all of the fast dependence of $\hat u$ is already included in the
term $\exp{\Big(\frac{i}{2}\int^{z}_{0}dz' [\omega^2
 \beta_2(z')+\frac{\omega^3}{3}\beta_3(z')]  \Big)}$ (see reference~\cite{gabtur1,gabtur2,lush2001a}).
This term is nothing more than an exact solution  of the linear part of Eq. $(\ref{nls0})$. In
first approximation one can neglect the slow dependence of $\hat \psi$ on $z$ in the interval
$mL\le z <(m+1)L$, i.e. one can replace $\hat \psi[\omega,z]$ by $\hat \psi[\omega,mL]$ in the
nonlinear term $R$ ($m$ is an arbitrary nonnegative integer number). This substitution allows to
rewrite Eq. $(\ref{nlsomega0})$ in the following form:
\begin{eqnarray} \label{nlsomega2}
 \hat \psi\big (\omega,(m+1) L\big)=\hat \psi(\omega,m L)+
 \nonumber \\
 i R\big(\hat \psi[\omega,mL],\omega,(m+1)L,mL\big)
 +
 O(\frac{L}{z_{nl}})^2.
\end{eqnarray}
The term $O(L/z_{nl})^2$ indicates the order of accuracy of this approximation. Eq.
$(\ref{nlsomega2})$ enables one to find $\hat \psi(\omega,(m+1) L)$ given $\hat \psi(\omega,m L)$.
Thus one can recover recover $u(t,z)$ using the definition of $\psi$. However for WDM simulation,
the accuracy $O(L/z_{nl})^2$ is not always sufficient. The next order approximation is obtained by
including the first order correction, $\hat \psi^{(1)}(\omega,mL),$ in the nonlinear term, $R$:
\begin{eqnarray}
 \hat \psi(\omega,(m+1) L)=\hat \psi(\omega,mL) \nonumber
\\
+i R(\hat \psi^{(1)}[\omega,z],\omega,z,mL) +O(\frac{L}{z_{nl}})^3,\label{psi1a}
\\
\hat \psi^{(1)}(\omega,z)\equiv \hat \psi(\omega,m L)+i R(\hat \psi[\omega,mL],\omega,z,mL).
\label{psi1b}
\end{eqnarray}

Equations $(\ref{Rdef2}),(\ref{psi1a}),$ and $(\ref{psi1b})$ form a closed set for the approximate
calculation of $\hat \psi(\omega,(m+1) L)$ given $\hat \psi(\omega,mL)$, where
$O(\frac{L}{z_{nl}})^3$ is the accuracy of the approximate solution which is controlled by the
appropriate choice of $L$.  The main obstacle in the numerical integration of Eqs.
$(\ref{Rdef2}),(\ref{psi1a}),$ and $(\ref{psi1b})$ is the computation of the integral term
$R\big(\hat v[\omega,z],\omega,z,mL\big)$ which generally requires $M\times N^3$ operations for
each iteration, where $N$ is the number of grid points in $\omega$ or $t$-space and $M$ is the
number of grid points for integration over $z$. Next one presents a very efficient numerical
algorithm for calculations $R\big(\hat v[\omega,z],\omega,z,mL\big)$.

In $t$-space Eq. $(\ref{Rdef2})$ becomes
\begin{eqnarray} \label{Rdef3}
\hat F^{-1}\Big (R\big(\hat v[\omega],\omega,z,mL\big) \Big )\nonumber\\
=\int^{z}_{mL}dz' \, c(z')
 {\bf G}^{(z')}\big (V^{(z')}(t,z')\big),
\end{eqnarray}
where $\hat F^{-1}$ is the inverse Fourier transform over $\omega$; $V^{(z)}(t,z) \equiv
|v^{(z)}(t, z)|^2v^{(z)}(t,z)$ and $\bf{ G}^{(z)}$ is the integral operator corresponding to the
multiplication operator $\hat{\bf G}^{(z)}\big (\hat\Psi^{(z)}(\omega,z)\big)\equiv
\exp{\Big(-\frac{i}{2}\int^{z}_{0}dz' [\omega^2
 \beta_2(z')+\frac{\omega^3}{3}\beta_3(z')]  \Big)}
\hat V^{(z)}(\omega,z)$ in the $\omega$-space. It follows from Eqs.
$(\ref{Rdef2}),(\ref{psi1a})-(\ref{Rdef3})$ that the numerical procedure for calculation of $R(\hat
A,\omega)$ requires the following eight steps:

(i) The Inverse Fourier Transform of $\hat v^{(z)}(\omega, m L)=\hat \psi(\omega, m L
)\exp{\Big(\frac{i}{2}\int^{z}_{0}dz' [\omega^2
 \beta_2(z')+\frac{\omega^3}{3}\beta_3(z')]  \Big)}$
for every value of $z$ $(mL< z \le (m+1)L).$

(ii) A calculation of $V^{(z)}(t,m L)$ from $v^{(z)}(t,m L)$.

(iii) The Forward Fourier Transform of $V^{(z)}(t,m L).$

(iv) A numerical integration (summation) of $c(z') \exp{\Big(-\frac{i}{2}\int^{z}_{0}dz' [\omega^2
 \beta_2(z')+\frac{\omega^3}{3}\beta_3(z')]  \Big)}\hat V^{(z')}(\omega)$ over
$z'$ (from $z'=mL$ to $z'=z$) for every values of $\omega$ and $z \ (mL < z\le(m+1)L).$ This
integration gives $\hat \psi^{(1)}(\omega,z)$ according to Eq. $(\ref{psi1b})$.

(v) The Inverse Fourier Transform of $\hat v^{(z)}(\omega, z)=\hat \psi^{(1)}(\omega, z
)\exp{\Big(\frac{i}{2}\int^{z}_{0}dz' [\omega^2
 \beta_2(z')+\frac{\omega^3}{3}\beta_3(z')]  \Big)}$
for every value of $z$, $mL< z \le (m+1)L$. (Note that in contrast with step (i) it is necessary to
take into account the dependence of $\hat \psi^{(1)}$ on $z$).

(vi)-(viii) These steps are similar to steps (ii)-(iv) except that the new value of $\hat
v^{(z)}(\omega, z)$ is used which was obtained in step (v).

The forward and inverse Fourier transforms can be performed with the fast Fourier transform (FFT)
which requires $N Log_2(N)$ numerical operations. Steps (i)-(iii) need only the value of
$\psi(t,mL).$ These steps can be performed independently and simultaneously in a network of $M$
central processor units (CPUs), shown schematically in Fig. 1. The number of CPUs, $M,$ coincides
with the number of grid points for integration over $z$. Thus the effective computational time
equals to time necessary to perform $2N Log_2(N)$ operations on complex numbers in one CPU. Below
to estimate effective computational time  one always refers to the number of numerical operations
in one CPU if all calculations can be implemented simultaneously in different CPUs without
communication between them.

The resulting values of $V^{(z)}(t,m L)$ (after step (iii)) are a set of $M$ vectors ${\bf a}_m, \
(m=1,2,\ldots, M)$ consisting of $N$ complex numbers each. Every vector ${\bf a}_m$ is stored in
the memory of the $m$th CPU (or in memory assigned to $m$th CPU in shared memory network). To
perform step (iv) one replaces these vectors by the new vectors ${\bf b}_m$: ${\bf
b}_m=\sum^m_{j=1} {\bf a}_j, \ (m=1,2,\ldots, M)$. Here a simple parallel algorithm is given. Note
that this algorithm can be improved but this improvement is outside the scope of this Letter. It
is assumed that $M$ is a power of 2: $M\equiv 2^{M_e}$, $M_e$ is an integer. The proposed
algorithm requires $M_e$ substeps. The vectors ${\bf b}^{(k)}_m, \ (m=1,2,\ldots, M)$ are results
of $k$th substep stored in memory. So that ${\bf b}^{(M_e)}_m={\bf b}_m$. The first substep is to
sum up every pair of vectors: ${\bf a}_{2m}+{\bf a}_{2m+1}$ to get ${\bf b}^{(1)}_1={\bf a}_1, \
{\bf b}^{(1)}_2= {\bf a}_1+{\bf a}_2, \ldots, \ {\bf b}^{(1)}_{M-1}={\bf a}_{M-1}, \
 {\bf b}^{(1)}_{M}={\bf a}_{M-1}+{\bf a}_M$. This summation requires $N$ operations.
 By induction one can see that after $k$ substeps, ${\bf b}^{(k)}_m=\sum^{m}_{j=1}{\bf a}_j$
 for $1\le m\le 2^k,$
 ${\bf b}^{(k)}_m=\sum^{m}_{j=2^k+1}{\bf a}_j$ for $2^k+1\le m\le 2^k+2^{k}, \ldots,$
 ${\bf b}^{(k)}_m=\sum^{m}_{j=2^{M_e}-2^k+1}{\bf a}_j$ for $2^{M_e}-2^k+1\le m\le 2^{M_e}.$
 Note that M vectors are now grouped in $M/2^k$ blocks with the appropriate summation inside each block.
 To perform the $k+1$th substep, it is necessary to double the block size. This
 can be done by adding the last
 element of each odd block to each element of next even block. To do this, one first creates
in memory  of $2^k$ copies of the last element of each odd block, which requires $k N$ operations
in a parallel CPU network. (A number of copies can be doubled by memory forking after each $N$
operations). To complete the $k+1$th substep, it is now enough to simultaneously add $2^k$ copies
to each element in the even block, requiring $N$ operations. The total number of operations for
step (iv) is $[1+2+\ldots M_e]N=M_e(M_e+1)/2$. Steps (v)-(viii) can be done in about $N[2
Log_2(N)+Log_2(M)]$ operations. (In step (viii) it is only necessary to calculate ${\bf b}_M$
requiring $NLog_2[M]$ operations). Thus the total number of operation for steps (i)-(viii) is:
\begin{eqnarray}\label {Ntot}
N[4 Log_2(N)+Log_2(M) +Log_2(M)
\nonumber \\
\times \big(Log_2(M)+1\big )/2]\sim N[4 Log_2(N)+\frac{Log_2(M)^2}{2}].
\end{eqnarray}

Direct solution of $(\ref{nls0})$ by a split-step method with the same accuracy (for the same size
of numerical step, $L/M,$ and the same number of points $N$ in $\omega$ space) requires $2M N
Log_2(N)$ operations. Comparing this with $(\ref{Ntot}),$ one can conclude that the proposed
parallel algorithm allows one to do numerical simulations with the same numerical accuracy about
$M/2$ times faster using a network of $M$ parallel CPUs. However the proposed algorithm is about 2
times slower if only one CPU is used.

Numerical simulations of the WDM system were performed using both the split-step method for NLS
$(\ref{nls0})$ and using the numerical algorithm given by Eqs. $(\ref{Rdef2}),(\ref{psi1a}),$ and $
(\ref{psi1b})$ to demonstrate the accuracy of the proposed numerical scheme. Simulations were
performed for 5 WDM channels (20 Gb/s per channel) over a typical transoceanic distance of $10^4 \,
km$. The channel spacing was $0.6 \, nm.$ The GVD periodically alternates between spans of standard
monomode fiber ($\beta_2^{(1)}=-20.0 \, ps^2/km,$ $\beta_3^{(1)}=0.1  \, ps^3/km,$
$\sigma_1=0.0013 \, (km \, mW)^{-1}$, length $L_1=40\, km$) and dispersion compensating fiber
($\beta_2^{(2)}=103.9 \, ps^2/km,$ $\beta_3^{(2)}=-0.3  \, ps^3/km,$ $\sigma_2=0.00405\, (km \,
mW)^{-1}$, length $\ L_2=-\beta_2^{(1)} L_1/\beta_2^{(2)}\, km$) so that the average GVD is zero.
 Fiber losses and
amplifiers were not considered. However they can be easily included in the coefficient $c(z).$ A
pseudo-random binary sequence of length 20 was used for every WDM channel. The boundary conditions
are periodic in time. Each binary ``1" was represented by an initially zero-chirp Gaussian pulse
(return to zero format) of $10ps$ width and peak power $|u|^2=1\, mW$ at the beginning $(z=0)$ of
the fiber line which is taken at the middle of standard monomode fiber span. The integration length
$L$ (see Eqs. $(\ref{Rdef2}),(\ref{psi1a}),(\ref{psi1b})$) is set to be equal to $(L_1+L_2)/4$;
$M=2^9;$ and $N=2^{11}.$ Fig. 2 shows the pulse power distribution (simultaneously in all 5
channels) after propagating $10^4 km$  obtained from both the split-step and the proposed parallel
algorithm. The differences in power distribution between these two simulations are less than 1$\%$
so the two curves are indistinguishable in Fig. 2. Numerical simulations were performed on usual
workstation without use of parallel computations. The objective of this numerical example is to
demonstrate the relative accuracy of numerical algorithm.  Hardware implementation of the parallel
simulation for numerical algorithm $(\ref{Rdef2}),(\ref{psi1a}),$ and $ (\ref{psi1b})$ is beyond
the scope of this Letter.

One can conclude that the proposed parallel numerical algorithm allows one to implement  numerical
simulations of Eq. $(\ref{nls1})$ about $M/2$ times faster than a direct numerical simulation of
that Eq. by the split-step method with the same accuracy. The absence of communications between
parallel CPUs during the computation of the FFT allows one to implement the proposed massive
parallel algorithm  on workstation clusters.


The author thanks M. Chertkov, I.R. Gabitov and S. Tretiak for helpful discussions.

Support was provided by the Department of Energy, under contract W-7405-ENG-36.

E-mail address: lushnikov@cnls.lanl.gov.




\newpage

\section*{Figure captions:}

~

\noindent Fig.1. A schematic of parallel computation algorithm and required number of numerical
steps. $FFT_1, \ FFT_2 , \ldots$ represent FFT in first, second etc. CPUs respectively.
Righ-hand-side schematically shows calculation of vectors ${\bf b}_m$ (see text).

\noindent Fig.2. Power distributions of 5 WDM channels after propagation of pseudorandom sequences
of Gaussian pulses over $10^4$ km.  Only small part of the total computational interval of $1000
ps$ is shown.

\newpage

\begin{figure*}
\epsfxsize=12.5cm \epsffile{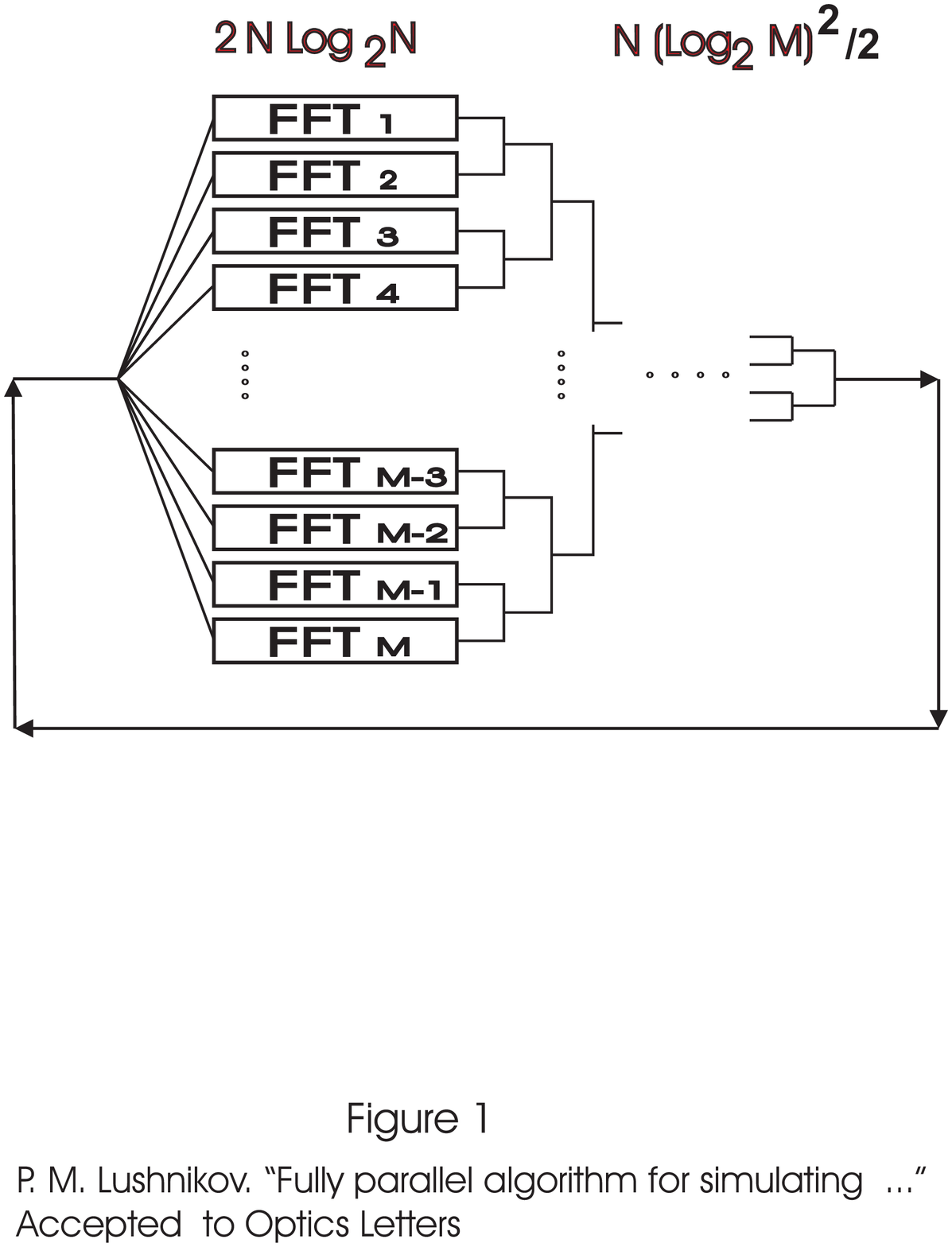}
\end{figure*}

\newpage

\begin{figure*}
\epsfxsize=15.5cm \epsffile{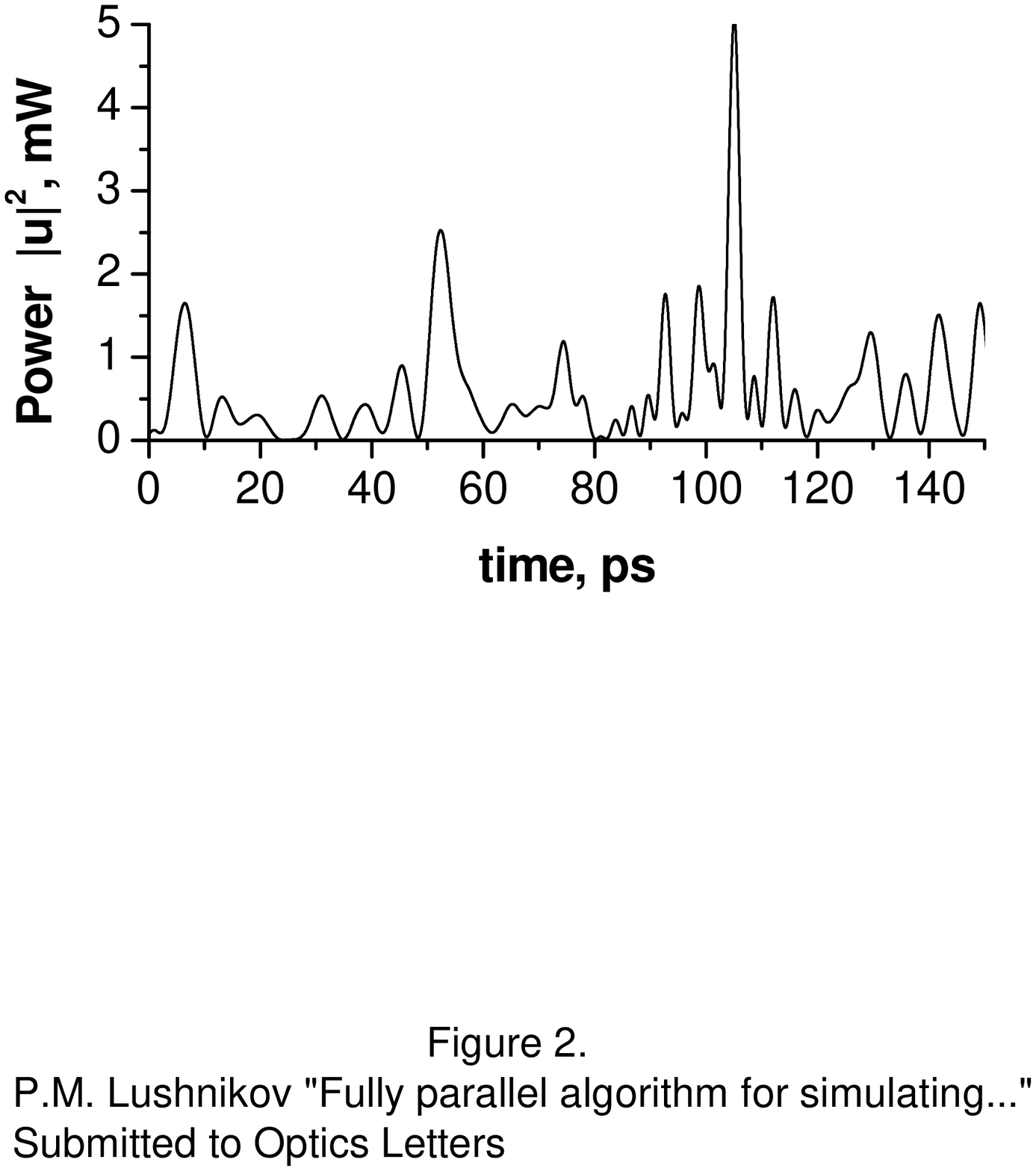}
\end{figure*}

\end{document}